\documentclass[12pt]{article}
\pdfoutput=1	
\usepackage[utf8]{inputenc}
\usepackage{array} 
\usepackage{amsmath,amssymb}
\usepackage{graphicx}
\usepackage[pdftex]{hyperref}
\usepackage{caption}
\usepackage{xcolor}
\DeclareGraphicsExtensions{.eps,.bmp,.wmf,.jpg}
\numberwithin{equation}{section}
\def\be{\begin{equation}}
\def\ee{\end{equation}}

\def\bea{\begin{eqnarray}}
\def\eea{\end{eqnarray}}

\title{Modified Gravity in the framework of holographic dark energy}
\author{L.N. Granda\thanks{luis.granda@correounivalle.edu.co} ,\, G. D. Rojas\thanks{german.dario.rojas@correounivalle.edu.co}\\{\it Departamento de Fisica, Universidad del Valle}\\{\it A.A. 25360, Cali, Colombia}}
\date{}
\begin{document}
\maketitle

\begin{abstract}
The modified gravity is considered in the framework of the holographic dark energy. An analysis of the autonomous system, the critical points and their stability is presented. Unlike the dark energy models based on $f(R)$, it is found that working in the holographic frame enriches the possibility of accelerated and matter type points for different cosmological scenarios, making viable trajectories of successful $f(R)$ models that are not allowed without the consideration of the holographic framework. The implications for the Hu-Sawicki model are analyzed.\\

\end{abstract}

\maketitle


\section{Introduction}

One of the alternatives to explain the dark energy issue (for review see \cite{NojiriOdintsov2007b}, \cite{Tsujikawa2010}, \cite{SotiriouFaraoni2010}, \cite{NojiriOdintsov2014}) is the large-distance modification of gravity, also called $f(R)$, where the late-time acceleration of the universe is associated to the modified (compared with General Relativity) response of the geometry of space-time to the presence of matter/energy. This is achieved by changing the form of the EH lagrangian by a general function of the Ricci scalar $R$, represented as $f(R)$.\\
 This alternative has been extensively studied and several proposals of modified gravity (exponential, power law, broken power law, logarithmic, hypergeometric, hyperbolic, stepped functions, reconstructed, etc.) have been presented to explain early and late time cosmic acceleration, and a combination of both (\cite{Ruzmaikina1970}, \cite{Starobinsky1980}, \cite{Barrow1983}, \cite{Schmidt2001}, \cite{Capozziello2002-2003}, \cite{Carroll2004}, \cite{NojiriOdintsov2003}, \cite{NojiriOdintsov2003a}, \cite{Capozziello2003}, \cite{Brookfield2006}, \cite{delaCruz2006}, \cite{ApplebyBattye2007}, \cite{HuSawicki2007}, \cite{NojiriOdintsov2006}, \cite{NojiriOdintsov2006a}, \cite{NojiriOdintsov2007}, \cite{NojiriOdintsov2007a}, \cite{NojiriOdintsov2008}, \cite{NojiriOdintsov2008a}, \cite{NojiriOdintsov2008b}, \cite{Cognola2008}, \cite{Miranda2009}, \cite{Starobinsky2007}, \cite{Deser2007}, \cite{Maggiore2014}, \cite{Harko2012}, \cite{Granda2014}, \cite{Tsujikawa2008}, \cite{Linder2009}, \cite{Bamba2010}, \cite{Yang2010}, \cite{Chen2015}). In general, the $f(R)$ models must fit the results of the observations \cite{JainTaylor2003}, \cite{Tegmark2004Seljak2005}, \cite{Eisenstein2005Blake2006}, \cite{Planck2015}, \cite{Perlmutter1999Riess1998Tonry2003Knop2003}, \cite{Abbott2017}, \cite{Abbott2017a}, \cite{Abbott2017b} and also accomplish several theoretical requirements (\cite{Sotiriou2008}, \cite{Amendola2007}), and describe correctly the different evolutionary stages of the universe. Also, the viable modified gravity models implement the so called \textit{Chamaleon Mechanism} (\cite{KhouryWeltman2004}, \cite{KhouryWeltman2004a}) through which they can pass the solar system tests.\\
On the other hand, the holographic principle (\cite{tHooft1993}, \cite{Susskind1995}, \cite{Fischler1998}, \cite{Horava2000}, \cite{Thomas2002}) was born motivated by the theoretical developments on quantum mechanics of black holes, and as an alternative to build a quantum theory of gravity. The energy coming from this principle (holographic energy density \cite{Li2004}) can be taken as a source of a non-constant vacuum energy. 
Karami and Khaledian \cite{KaramiKhaledian2011}, and Majumder \cite{Majumder2013} proposed the reconstruction of several $f(R)$ models having as source of them the energy coming from the holographic and agegraphic principle. Rezaei et. al \cite{Sola2019} presented several power law models of DE that include powers of the Hubble  parameter and its first derivative (including the holographic IR cut-off proposed by Granda and Oliveros \cite{GrandaOliveros2009}), and studied the behavior of $H(z)$ and the $EoS$ for each one.
The $\Lambda CDM$ model (or concordance model) is the one that better describes the behavior of the universe since the radiation era to the present. However, it presents the problem of fine tuning and the presence of a constant vacuum energy density whose origin is not clear. In this way, the source (or sources) of dark energy (DE)  that dominates the current universe (about $72 \%$) has not been yet identified, giving place to theoretical proposals describing the origin and behavior of this sector.\\
In this paper we suggest the possibility that the dark energy has two sources, one associated to $f(R)$ modified gravity and the other coming from a holographic energy density. The holographic principle by itself has universal character and, up to now, there has not been any observational or theoretical development that demonstrates the unfeasibility of the holographic principle. An argument in favor of considering these two sources, is that while $f(R)$ is motivated by the possibility that General Relativity does not provide the correct description of gravity at cosmological scales, the holographic principle is motivated by a very different problem. In fact, the holographic principle was proposed as a possibility to address the microscopic nature of gravity (or equivalently the quantum nature of vacuum). Also, because the theoretical developments on black hole quantum mechanics and thermodynamics, and the AdS/CFT correspondence, the holographic principle has been gaining ground as a fundamental piece to understand the nature of the universe. Though the holographic principle implies a unification of gravity, matter and  quantum mechanics, as long as there is no theory of quantum gravity, it is valid to assume that the theory of quantum gravity may imply a, yet unknown, geometrical modification of the standard gravitation model. Therefore, a holographic principle does not enter in conflict with modified gravity, but rather from the perspective of quantum gravity they could complement each other. Then if some modified gravity model is the correct one for the description of gravitational phenomena at local and cosmological scales, them the holographic principle does not disappear but must be constructed within the framework of such theory. On the other hand, not because dark energy has two sources the universe becomes more phantom, on the contrary, as will be shown by the results below, with the inclusion of the holographic source the behavior of the equation of state  improves, covering more regions with viable values than in the standard $f(R)$, and being consistent with current observations.\\
We consider the holographic density proposed in \cite{GrandaOliveros2009} in the framework of  modified gravity, where a coupling appears to the field $F$ $\Big(\rho_\Lambda=\frac{3F}{\kappa^2}\big(\alpha \dot{H}+\beta H^2 \big)$, where $F=\frac{df(R)}{dR}\Big)$. The coupling between the IR cut-off (that defines the holographic component of dark energy) and $F$ can be seen as something natural because when conformally transforming the $f(R)$ models to the Einstein frame, there appears a non-minimal coupling between the matter/energy sources (including the holographic energy) and an equivalent scalar field $\phi$ constructed from $F$. It produces an open continuity equation ($\neq 0$) for the pure (non-coupled) holographic fluid,  $\tilde{\rho}_\Lambda=\frac{3}{\kappa^2}\big(\alpha \dot{H}+\beta H^2 \big)$, equation that is only equal to zero in the case of General Relativity ($f(R)\rightarrow R$). In other words, the continuity equation has more dynamical character as it includes the coupling to the scalar field constructed form $F(R)$.\\
The characteristics of the autonomous system associated to the proposal are studied, and the critical points are described. 
It is possible to see that in the $m$ vs. $r$ plane, three of these critical points change their stability characteristics compared to those associated to DE coming from $f(R)$. As can be seen in \cite{Amendola2007}, there is a de Sitter point located at $r=-2$ that is stable in the interval $0<m<1$. In the holographic frame, this point is located at any value of r depending on the value of $\alpha$ ($\alpha\neq -1$) and becomes stable for a wider range of values of m. In \cite{Amendola2007}, there is an accelerated type point that has several characteristics: is de Sitter type (with marginal stability) for $m=1$, quintessence for $m<-\frac{\sqrt{3}+1}{2}$ and $\frac{\sqrt{3}-1}{2}<m<1$, and phantom for $-\frac{1}{2}<m<0$ and $m>1$ with $\omega_{eff}\notin{(-7.6,-1.07)}$ . In the present proposal this point can become de Sitter attractor in $m=1$, can achieve de Sitter scenarios for a wider range of values of $m$, can lead to quintessence eras, and becomes phantom for several intervals of $m$ that depend on the holographic parameters $\alpha$ and $\beta$ but now $\omega_{eff} \in (-\infty,-1)$. In \cite{Amendola2007} the matter point cannot be reached by cosmological trajectories with $m\rightarrow 0^-$ due to the positive divergence of one of the eigenvalues of the transformation matrix from the unperturbed to the perturbed GDE sector, while with the introduction of $\rho_\Lambda$, trajectories with $m\rightarrow0^-$  are now possible, giving place to long enough matter eras letting in this way the formation of structures in the universe and posterior evolution to other cosmological scenarios.\\
The results are applied to the Hu-Sawicki \cite{HuSawicki2007} $f(R)$ model for some specific values of the parameters of the model and the holographic parameters ($\alpha$ and $\beta$), and find that several trajectories that were not allowed, are now viable with the  introduction of  $\rho_\Lambda$ as vacuum energy.\\
This paper is organized as follows: In section II the general field equations for $f(R)$ models including holographic vacuum energy are written. In section III the autonomous system is presented and the corresponding critical points are found and the stability conditions for some interesting critical points are obtained. In section IV we compare the results of Amendola et al. \cite{Amendola2007} with the results of the present proposal, and apply the results to two trajectories of the Hu and Sawicki's modified gravity model. In section V some discussion is presented. 
\section{Field equations}

The modified gravity formalism is based on the Lagrangian and field equations of general function of curvature $f(R)$ and its results must respect the restrictions set by  local and large scale observations where the Einstein's theory and $\Lambda CDM$ give, so far, the most appropriate description.\\

\noindent The general action for $f(R)$ models is\\

\be\label{action}
S = \int d^4x \sqrt{-g} \left[  \frac{1}{2\kappa^2}f(R)+{\cal L}_m(g_{\mu\nu},\psi)\right],
\ee
where the space-time is four dimensional, any Greek index runs from 0 to 3, where 0 corresponds to the time dimension, and 1, 2 and 3 to the spatial ones; $\kappa^2$ is the gravitational constant ($\kappa^2=8\pi G$), $g$ is the determinant of the metric $g_{\mu\nu}$, and $\mathcal{L}_m$ is the lagrangian density for the matter sector that can be baryonic, dark matter, or any kind of exotic type of matter/energy. \\
By varying the action with respect to the metric $g_{\mu\nu}$, we find the field equations

\begin{equation}
    \label{fieldequation1}
    f'(R)R_{\mu\nu}-\frac{1}{2}f(R)g_{\mu\nu}-\bigg(\nabla_\nu\nabla_\mu-g_{\mu\nu} \square \bigg)f'(R)=\kappa^2 T_{\mu\nu}^{(m)}
\end{equation}\\
where $\square\equiv\nabla^\sigma\nabla_\sigma$ is the covariant d'Alembertian operator, and $$T_{\mu\nu}^{(m)}\equiv -\frac{2\kappa^2}{\sqrt{-g}}\frac{\delta\mathcal{L}_m}{\delta g^{\mu\nu}}$$ is the energy -momentum tensor for the matter sector.\\

The equation (\ref{fieldequation1}) can also be written as:

\begin{multline}
    \label{fieldequationeinstein}
    R_{\mu\nu}-\frac{1}{2}Rg_{\mu\nu}={\kappa^2}\tilde{T}_{\mu\nu}^{(m)}\\+\kappa^2\Bigg[\frac{1}{\kappa^2f'(R)}\bigg[\frac{f(R)-Rf'(R)}{2}g_{\mu\nu}+(\nabla_\nu\nabla_\mu-g_{\mu\nu}\square)f'(R)\bigg]\Bigg]
\end{multline}\\
where the last one term at right can be associated to the effective energy-momentum tensor coming from $f(R)$. The trace equation reads

\begin{equation}
    \label{traceeqn}
    f'(R)R-2f(R)+3\square f'(R)=\kappa^2T\hspace{0.3
    cm}.
\end{equation}\vspace{0.01cm}

\noindent By using the $FLRW$ metric, assuming a vanishing spatial curvature $K$ (which is supported by current observations) and taking the usual form of the energy-momentum tensor for a perfect fluid, the time and space-like components of the field equations are respectively

\begin{equation}
    \label{timelikeeqn}
    3H^2f'(R)=\kappa^2\rho +\frac{1}{2}\bigg [Rf'(R)-f(R)\bigg]-3H\dot{R}f''(R)
\end{equation}

and

\be\label{spacelikeeqn}
\begin{aligned}
(2\dot{H}+3H^2)f'(R)=&-\bigg[\kappa^2p+2H\dot{R}f''(R)+\frac{f(R)-Rf'(R)}{2}\\&+\dot{R}^2f'''(R)+\ddot{R}f''(R)\bigg],
\end{aligned}
\ee

\noindent where a dot over a quantity means derivative respect to time, $\frac{\dot{a}}{a}\equiv H$ is the Hubble parameter, $\dot{R}f''(R)=\partial_0f'(R)$, and $f''(R)$ and $f'''(R)$ are the second and third derivatives of $f(R)$ with respect to $R$. The trace equation becomes

\begin{equation}
    \label{traceeqnFLRW}
    f'(R)R-2f(R)+3\square f'(R)=\kappa^2(3p_m-\rho_m)
\hspace{0.3cm}.
\end{equation}

\noindent Taking into account the holographic principle, the density $\rho$ and pressure $p$ in equations(\ref{timelikeeqn}) and (\ref{spacelikeeqn}) include the corresponding holographic density $\rho_{\Lambda}$ and holographic pressure $p_{\Lambda}$, and these equations can be rewritten as

\begin{equation}
    \label{timelikeeqnEins}
    H^2=\frac{\kappa^2}{3}\big(\tilde{\rho}_m+\tilde{\rho}_\Lambda+\rho_f\big)
\end{equation}
and
\begin{equation}
    \label{spacelikeeqnEins}
    2\dot H+3H^2=-\kappa^2\big(\tilde{p}_m+\tilde{p}_\Lambda+p_f\big)
\end{equation}\\
where $\tilde{\rho}_m\equiv\frac{\rho_m}{f'(R)}$\hspace{0.15cm},\hspace{0.15cm}$\tilde{p}_m\equiv \frac{p_m}{f'(R)}$\hspace{0.15cm},\hspace{0.15cm}$\tilde{\rho}_\Lambda\equiv\frac{\rho_\Lambda}{f'(R)}$\hspace{0.15cm},\hspace{0.15cm},\hspace{0.15cm}$\tilde{p}_\Lambda\equiv\frac{p_\Lambda}{f'(R)}$\hspace{0.15cm}, with

\be\label{rhol}
\rho_\Lambda=\frac{3F}{\kappa^2}\big(\alpha\dot{H}+\beta H^2 \big),
\ee

\begin{equation}\label{rhof}
 \rho_f=\frac{1}{\kappa^2f'(R)}\bigg[\frac{Rf'(R)-f(R)}{2}-3H\dot{R}f''(R)\bigg]
\end{equation}
and
\be\label{pf}
p_f=\frac{1}{\kappa^2f'(R)}\bigg[2H\dot{R}f''(R)+\frac{f(R)-Rf'(R)}{2}+\\(\dot{R})^2f'''(R)+\ddot{R}f''(R)\bigg].
\ee\\
If one writes the Lagrangian density for modified gravity as
$$f(R)=1+\tilde{f}(R),$$ then the consistency with high redshift universe demands that $\tilde{f}(R)<<R$ and $|\tilde{f},_{R}|<<1$. This last restriction means that $F(R)=1+\tilde{f},_{R}$ must be very close to 1 (which should be satisfied by any viable modified gravity model), and therefore the expression (\ref{rhol}) maintains very close to the holographic density proposed for the Einstein gravity. 
The equation (\ref{timelikeeqnEins}) can be written as

\begin{equation}
    \label{timelikeeqnwithF}
    1=\frac{R}{6H^2}-\frac{f}{6H^2F}-\frac{\dot{F}}{HF}+\frac{\kappa^2\rho^{(m)}}{3H^2F}+\alpha+\beta\frac{\dot{H}}{H^2}
\end{equation}\\
which, defining the dynamical variables 

\begin{equation}
    \label{dynamicalvariables}
    x\equiv-\frac{\dot{F}}{HF} \hspace{0.45cm}y\equiv-\frac{f}{6H^2F}\hspace{0.45cm}z\equiv\frac{R}{6H^2}=2+\frac{\dot{H}}{H^2}
\end{equation}\\[-0.5ex] 
takes the form

\begin{equation}
    \label{timelikeeqnwithdynamical}
    1=x+y+(1+\beta)z+\alpha-2\beta+\Omega_m\hspace{0.3cm}.
\end{equation}

\noindent As can be seen, the dynamical variables and the holographic parameters $\alpha$ and $\beta$ allow to write the DE density parameter coming from $f(R)$ and $\rho_\Lambda$ as:

\begin{equation}
    \label{DEdensityparameter}
    \Omega_{DE}\equiv (x+y+z)+\bigg(\alpha+\beta\frac{\dot{H}}{H^2}\bigg)=x+y+(1+\beta)z+\alpha-2\beta
\end{equation}\\
where $\Omega_f\equiv x+y+z$\hspace{0.3cm}and\hspace{0.3cm}$\Omega_\Lambda\equiv\alpha+\beta\frac{\dot{H}}{H^2}$.
\section{Autonomous system and critical points}

After some algebra and defining the parameter $m\equiv\frac{RF'}{F}=\frac{dLnF}{dLnR}$, it is possible to find the equations for the critical points

\be\label{59eq}
\begin{aligned}
    \frac{dx}{dN}=&x^2+\bigg(\frac{\beta}{m}+\beta -1\bigg)xz+\big(\alpha-2\beta\big)x\\& -3y+\big(\beta-2\alpha-1)z+\alpha-2\beta-1
\end{aligned}    
\ee
\begin{equation}
    \label{60eq}
    \frac{dy}{dN}=xy+\frac{xz}{m}-2y(z-2)
\end{equation}
\begin{equation}
    \label{61eq}
    \frac{dz}{dN}=-\frac{xz}{m}-2z(z-2)
\end{equation}\\
where $N$ is the e-folding variable defined as $N\equiv Ln \hspace{1mm}a$.\\ 

\noindent The critical points of this system ($(x_c, y_c, z_c)$) can be found by equating to zero the above equations.\\

\noindent The matter content of the universe can be found from (\ref{timelikeeqnwithdynamical}) as

\begin{equation}
    \label{matterdensityparameter}
    \Omega_m=1-x-y-(1+\beta)z-\alpha+2\beta\hspace{0.3cm}.
\end{equation}\\
On the other hand, replacing equation (\ref{timelikeeqnEins}) in (\ref{spacelikeeqnEins}) acceleration equation becomes

\begin{equation}
    \label{63eq}
    \frac{\ddot{a}}{a}=-\frac{\kappa^2}{6}\tilde{\rho}(1+3w_{eff})
\end{equation}\\
where $\tilde{\rho}=\tilde{\rho}_m+\tilde{\rho}_f+\tilde{rho}_{\Lambda}$ and the \textit{EoS} is given by $w_{eff}=\frac{\tilde{p}}{\tilde{\rho}}=-1-\frac{2}{3}\frac{\dot{H}}{H^2}=-\frac{1}{3}(2z-1)$, being $z$ the dynamical variable already defined.\\

Then, the critical points of the model with the respective values of $\Omega_m$ and $w_{eff}$ are\\

\be\label{cp1}
P_1=(0, -1-\alpha, 2),\;\;\; \Omega_{m_{P1}}=0,\;\;\; w_{eff_{P1}}=-1.
\ee
\be\label{cp2}
P_2=(-1, 0, 0),\;\;\; \Omega_{m_{P2}}=2-\alpha+2\beta,\;\;\; w_{eff_{P2}}=\frac{1}{3}.
\ee
\be\label{cp3}
P_3=(1-\alpha+2\beta, 0, 0),\;\;\; \Omega_{m_{P3}}=0,\;\;\; w_{eff_{P3}}=\frac{1}{3}.
\ee
\be\label{cp4}
P_4=(-4, 5-\alpha+2\beta,0),\;\;\; \Omega_{m_{P4}}=0,\;\;\; w_{eff_{P4}}=\frac{1}{3}.
\ee
\be\label{cp5}
\begin{aligned}
P_5=&\bigg(\frac{3m}{1+m},-\frac{1+4m}{2(1+m)^2},\frac{1+4m}{2+2m}\bigg),\\
&\Omega_{m_{P_5}}=\frac{2-2\alpha-2m^{2}(4+\alpha)+3\beta+m(-3-4\alpha+3\beta)}{2(1+m)^{2}},\\
& w_{eff_{P5}}=-\frac{m}{1+m}
\end{aligned}
\ee

\be\label{pc6}
\begin{aligned}
 P_6=&\bigg(\frac{2m(1-m-\alpha-m\alpha)}{m+2m^2-\beta-m\beta},\frac{1-4m-\alpha+2\beta}{m+2m^2-\beta-m\beta}, \frac{(1+m)(-1+4m+\alpha-2\beta)}{m+2m^2-\beta-m\beta}\bigg)\\
&\Omega_{m_{P_6}}=0,\\&  
w_{eff_{P6}}=\frac{1}{3}\bigg[1-\frac{2(1+m)(-1+4m+\alpha -2\beta)}{m+2m^2-\beta-m\beta}\bigg]. 
\end{aligned}
\ee

\noindent {\bf Eigenvalues and stability conditions.}\\

\noindent In some interesting critical points we calculate the eigenvalues of the matrix that transforms the system from an unperturbed to a perturbed state of DE coming from $f(R)$. The value and sign of these eigenvalues determine the evolution of perturbations (at first order) made on the modified sector\\
As can be seen, $P_1$ is a de Sitter type point whose corresponding eigenvalues are:\\

\be\label{EVP1}
EVP_1:\{-3, -3m+m\alpha+2\beta \pm\frac{1}{2m}[4(-4m+4m^2+4m\alpha+4m^2\alpha)\\+(-3m+m\alpha+2\beta)^2]^{1/2}\}
\ee    
Defining the parameter $r\equiv -\frac{R F}{f}=\frac{z}{y}$, it is possible to see that in $P_1$, $r=-\frac{2}{1+\alpha}$, and the stability in this point is achieved if\\

\hspace{1.3cm}$m<-1$\hspace{0.3cm} and \hspace{0.3cm} $\alpha<\frac{1-m}{1+m}$\hspace{0.2cm}
and \begin{equation} \beta\geq\frac{1}{2}(3m-\alpha m)+2\sqrt{m-\alpha m-m^2-\alpha m^2}\hspace{0.2cm},\end{equation}\\

\hspace{0.7cm}or\hspace{0.5cm}$-1<m<0$\hspace{0.3cm}and\hspace{0.3cm}$\alpha>\frac{1-m}{1+m}$\hspace{0.3cm}and\begin{equation}\beta\geq\frac{1}{2}(3m-\alpha m)+2\sqrt{m-\alpha m-m^2-\alpha m^2}\hspace{0.2cm},\end{equation}\\

\hspace{0.7cm}or\hspace{0.5cm}$m>0$\hspace{0.5cm}and\hspace{0.5cm}$\alpha<\frac{1-m}{1+m}$\hspace{0.5cm}and\begin{equation} \hspace{0.5cm}\beta\leq \frac{1}{2}(3m-\alpha m)-2\sqrt{m-\alpha m - m^2 - \alpha m^2}\hspace{0.2cm},\end{equation}\\
otherwise it is repulsive (unstable) or transitory (saddle) point.\\

\noindent Another point that can lead to accelerated scenarios is $P_6$. As can be seen, this point can be of Quintessence, de Sitter or Phantom nature depending on the values of $m$, $\alpha$ and $\beta$ ($w_{eff_{P6}}$ depends on these parameters). Its eigenvalues are
\be\label{EVP6}
\begin{aligned}
EVP_6:&\Big\{\frac{2(1+m')(1+m)(1-m-\alpha-m\alpha)}{m+2m^2-\beta-m\beta},\\&\frac{2-2\alpha-2m^2(4+\alpha)+3\beta+m(-3-4\alpha+3\beta)}{m+2m^2-\beta-m\beta},\\&\frac{1-4m-\alpha+2\beta}{m}\Big\}.
\end{aligned}
\ee

\noindent The stability of this point depends on the values of the holographic parameters $\alpha$ and $\beta$, $m$ and $m'\equiv\frac{dm}{dr}$ where $r$ is the parameter already defined.\\

\noindent The critical point $P_5$ can have a matter type behaviour if $m\rightarrow 0$ ($w_{eff_{P5}}\rightarrow 0$) and can have matter dominance depending on the values of  $m$, $\alpha$ and $\beta$.\\\\
Its eigenvalues are:
 \begin{equation}EV P_5:\Big\{3(1+m')\hspace{0.2cm},\hspace{0.2cm}\frac{ 2m^2\alpha+\beta+m(2\alpha+\beta-3)\pm\sqrt{A}}{4m(1+m)}\Big\}
 \end{equation}
 where 
 $A=4m^4(8+\alpha)^2+\beta^2+2m[\beta^2-15\beta+2\alpha(4+\beta)-8]+4m^3[40+2\alpha^2-24\beta+\alpha(33+\beta)]+m^2[4\alpha^2-126\beta+\beta^2+\alpha(84+8\beta)-31]$\hspace{0.3cm}.\\

\noindent In the limit $\mid m \mid\ll1$, where $\omega_{eff_{P_5}}\rightarrow0$ corresponding to the $EoS$ of matter, the eigenvalues approximately reduce to:
\begin{equation}\label{EVP5m0} 
EV P_5 (\mid m\mid\ll1):\hspace{0.2cm}\Big\{3(1+m')\hspace{0.2cm},\hspace{0.2cm}\frac{\beta}{2m}\hspace{0.2cm},\hspace{0.2cm}\frac{2\alpha+\beta-3}{4}\Big\}.\end{equation}

\noindent If $m\rightarrow0^-$ and $\beta>0$, the second eigenvalue of $P_5\rightarrow-\infty$ but the third eigenvalue can be modulated with the values of the holographic parameters $\alpha$ and $\beta$. If $m'<-1$ and $\frac{2\alpha+\beta}{4}>\frac{3}{4}$, $P_5$ is a saddle point with one non-divergent positive eigenvalue giving place to a saddle point that can lead to a long enough matter era. The same situation occurs if $m'>-1$ no matter the values of $\alpha$ and $\beta$ as long as they do not diverge. If $m\rightarrow0^+$ and $\beta<0$, the second eigenvalue of $P_5$ is negative divergent and the achievement of a suitable (long enough) saddle matter point is driven by $m'$ ($m'>-1$ or $m'<-1$) in the first eigenvalue, and $\alpha$ and $\beta$ $\big(2\alpha-3>\mid\beta\mid$\hspace{0.2cm}or\hspace{0.2cm} $2\alpha-3<\mid\beta\mid\big)$ in the third eigenvalue.\\ 

\noindent {\bf Comparison and Analysis of results.}\\

\noindent The critical points $P_1$, $P_6$ and $P_5$, and their respective eigenvalues out of the holographic frame (just with $f(R)$ as source of DE) \cite{Amendola2007} are:\\

   $\tilde{P}_1=(0, -1, 2)$\hspace{0.2cm},\hspace{0.2cm} $\Omega_{m_{\tilde{P}_1}}=0$ \hspace{0.2cm},\hspace{0.2cm}  $w_{eff_{\tilde{P}_1}}=-1$ \hspace{0.2cm}.\\[0.3ex]
   
    $\tilde{P}_6= \bigg(\frac{2m(1-m)}{m(1+2m)}, \frac{1-4m}{m(1+2m)}, \frac{(1+m)(4m-1)}{m(1+2m)}\bigg)$\hspace{0.2cm},
    
    $\Omega_{m_{\tilde{P}_6}}=0$\hspace{0.2cm},\hspace{0.2cm}$w_{eff_{\tilde{P}_6}}=\frac{2-5m-6m^2}{3m(1+2m)}$\hspace{0.2cm}.\\[0.3ex]
    
    $\tilde{P}_5=\Big(\frac{3m}{1+m},-\frac{1+4m}{2(1+m)^2},\frac{1+4m}{2(1+m)}\Big)$\hspace{0.2cm},
    
    $\Omega_{m_{\tilde{P}_5}}=1-\frac{m(7+10m)}{2(1+m)^2}$\hspace{0.2cm},\hspace{0.2cm}$w_{eff_{\tilde{P}_5}}=-\frac{m}{1+m}$\hspace{0.2cm}.\\
with eigenvalues
\be\nonumber
EV \tilde{P}_1:\hspace{0.1cm} \Big\{-3\hspace{0.2cm},\hspace{0.2cm}-\frac{3}{2}\pm\frac{\sqrt{25-\frac{16}{m}}}{2}\Big\}.
\ee
\be\nonumber   
EV \tilde{P}_6:\hspace{0.2cm}\Big\{\frac{1-4m}{m}\hspace{0.2cm},\hspace{0.2cm}\frac{2-3m-8m^2}{m(1+2m)}\hspace{0.2cm},\hspace{0.2cm}\frac{2(1+m')(1-m^2)}{m(1+2m)}\Big\}.
\ee
\begin{equation}\label{evs}
EV \tilde{P}_5:\hspace{0.2cm}  \Big\{3(1+m')\hspace{0.2cm},\hspace{0.2cm}\frac{-3m\pm\sqrt{m(256m^3+160m^2-31m-16}}{4m(m+1)}\Big\}.
\end{equation}

\noindent As can be seen, the de Sitter type point $\tilde{P}_1$ in the $m$ vs. $r$ plane is always located at $r=-2$ and becomes stable if $0<m<1$\hspace{0.1cm}. On the other hand, in the framework of holographic dark energy, the de Sitter point $P_1$ in the $m$ vs. $r$ plane is located at $r=-\frac{2}{1+\alpha}$ and stable under certain conditions on $\alpha$ and $\beta$ which depend on the value of $m$, but now for any $m$ excluding the values $\{-1,0\}$. It shows that the introduction of $\rho_\Lambda$ gives versatility to the point $P_1$ (compared to $\tilde{P}_1$) because in the $m$ vs. $r$ plane it can be located in a wider range of values of $r$ and can be stable not only in $0<m<1$. It gives the possibility that some trajectories of several viable $f(R)$ models can have de Sitter attractors in values of $m$ and $r$ which do not give place to de Sitter stable points out of the holographic frame.\\

\noindent In the case of $\tilde{P}_6$, it can be seen that if $m=1$, $\omega_{eff_{\tilde{P}_6}}=-1$ which corresponds to a de Sitter scenario. Analyzing the eigenvalues of $\tilde{P}_6$ with this value of $m$, two of them are negative, while the other one is zero, which corresponds to a marginal stability. In order to have certainty of the behaviour (stable or saddle) of this point, it is necessary to make use of the central manifold mechanism. However, in the frame of the holographic principle this point can be an attractor with suitable values of the holographic parameters $\alpha$ and $\beta$ as can be seen in $EV P_6$. The condition $m=1$ associated to the de Sitter behaviour of this point, can be integrated to give a specific solution of $f(R)$:
\begin{equation}\begin{split}m&=\frac{R F'}{F}=1\hspace{0.3cm} \Rightarrow\hspace{0.3cm} F'=\frac{dF}{dR}=\frac{F}{R}\hspace{0.3cm}\\&\Rightarrow\hspace{0.3cm}F\propto R\hspace{0.3cm} \Rightarrow\hspace{0.3cm}f(R)\propto R^2\hspace{0.3cm},\end{split}\end{equation}\\
which corresponds to the Starobinsky's model for inflation \cite{Starobinsky1980} when $\frac{R}{R^2} \rightarrow 0$.\\
The line $m=0$ in the $m$ vs $r$ plane corresponds to the $\Lambda CDM$ model. It is possible to see that if $m\rightarrow0$, $\tilde{P}_6$ becomes a singular point, and $\omega_{eff_{\tilde{P}_6}}\rightarrow \pm \infty$. However, taking into account $\rho_\Lambda$, $\Tilde{P}_6$ turns into $P_6$, which with $\alpha=1$ and $\beta \neq 0$ is no longer singular, $\omega_{eff_{P_6}}=-1$ and the $\Lambda CDM$ model is achieved. Also, if $\alpha\approx1$ and $\beta \neq 0$, $P_6$ is non-singular and quasi-de Sitter. \\
Another advantage of working in the holographic frame is that if $\alpha=\frac{1-m}{1+m}$\hspace{0.2cm}then\hspace{0.1cm} $\omega_{eff_{P_6}}=-1$, achieving in this way a general de Sitter solution (for any value of $m$ except $m=-1$) in $P_6$ which is not possible just with $f(R)$. Also, if
\begin{equation*}\beta=\frac{\frac{1}{3}(6m^2-17m-20 \alpha m-20\alpha+20)}{1+m}\end{equation*}\begin{equation}\text{or}\hspace{0.6cm}\beta=\frac{\frac{1}{3}(6m^2+23m+20 \alpha m+20\alpha-20)}{1+m}\hspace{0.3cm},\end{equation}\\
no matter the value of $\alpha$, $\omega_{eff_{P_6}}=-1.1$\hspace{0.2cm}and\hspace{0.2cm}$\omega_{eff{P_6}}=-0.9$ respectively, for any value of $m$ (except $m=-1$) in $P_6$, giving quasi-de Sitter solutions which are allowed for the universe according to the current observations.\\

\noindent The eigenvalues in $P_6$ for the de Sitter solution with  $\alpha=\frac{1-m}{1+m}$ are:
\begin{equation} EV P_6 (de Sitter):\hspace{0.1cm}\Big\{ 0\hspace{0.1cm},\hspace{0.1cm}-5+\frac{8}{1+m}+\frac{2\beta(m-3)+6m}{\beta+(\beta-1)m-2m^2}\hspace{0.1cm}, 2\Big(\frac{1}{1+m}+\frac{\beta}{m}-2\Big)\Big\},
\end{equation}
so if 
 \begin{multline}\hspace{1.9cm}m<-1 \hspace{0.2cm}\text{and}\hspace{0.2cm}\beta>\frac{m+2m^2}{1+m}\\
    \text{or}\hspace{0.2cm} -1<m<0 \hspace{0.2cm}\text{and}\hspace{0.2cm} \beta>\frac{2m^2+m}{1+m}\\
\text{or}\hspace{0.2cm}0<m\leq 1\hspace{0.2cm}\text{and}\hspace{0.2cm} \beta<\frac{10m^3+5m^2+3m}{3m^2+6m+3}\\\text{or}\hspace{0.2cm}m>1 \hspace{0.2cm}\text{and}\hspace{0.2cm} \beta<\frac{2m^2+m}{1+
m}\hspace{0.8cm},\end{multline}\\
the second and third eigenvalues are negative so the general de Sitter solution is marginally stable and eventually can lead to de Sitter attractors.\\

\noindent If the dark energy sector is associated only to $f(R)$, the shape of the $EoS$ vs. $m$ for $\tilde{P}_6$ is shown in Fig. 1.\\
\begin{figure}
\centering
\includegraphics[scale=0.6]{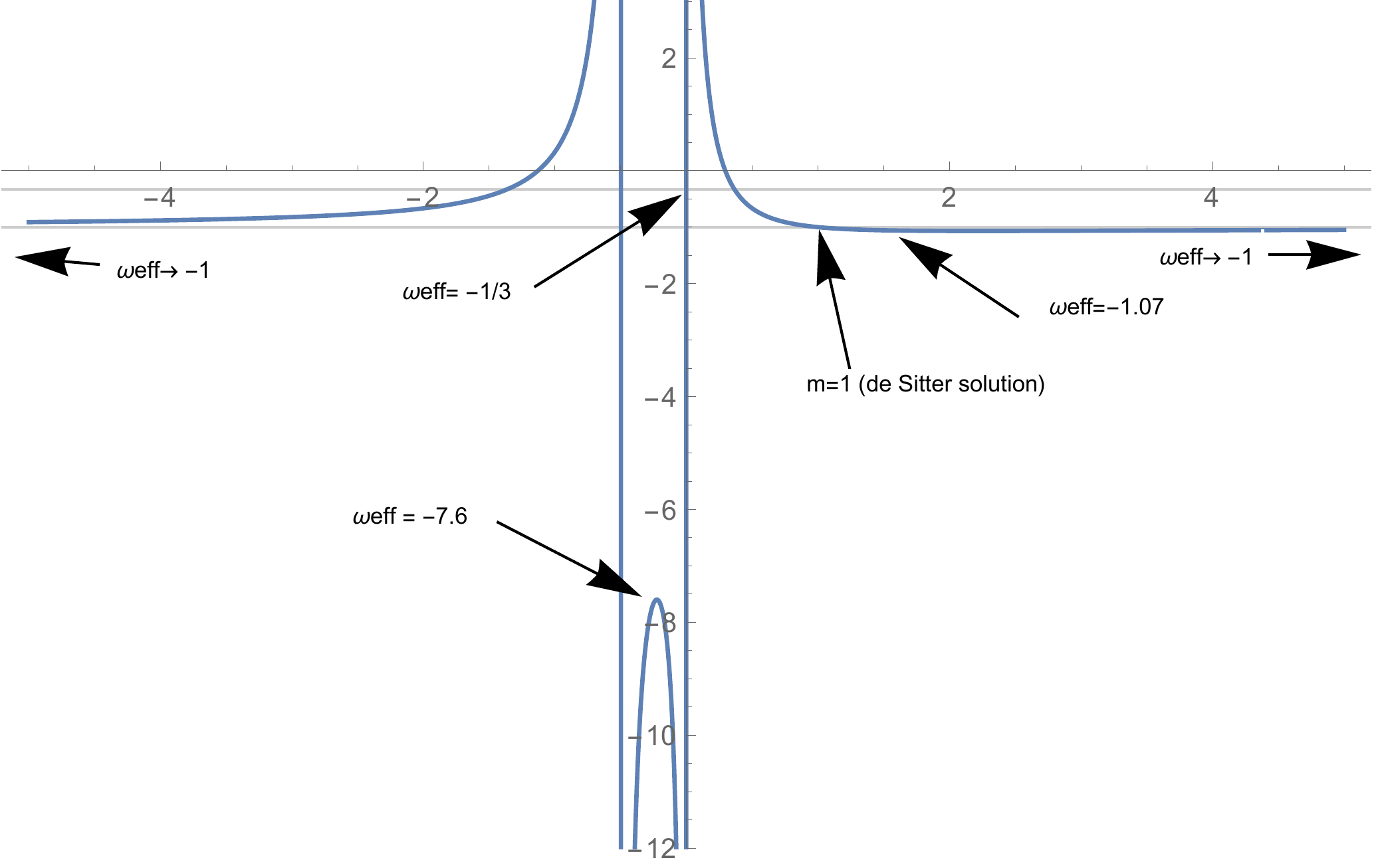}
\caption{The allowed regions for $w_{eff\tilde{P}_6}$ in terms of $m$.}
\end{figure}
\noindent But if it is viewed in the framework of the holographic principle, it takes the form shown in Fig. 2 for $\alpha=\frac{1}{3}$ and $\beta=\frac{1}{2}$ and in Fig. 3 for  $\alpha=3$ and $\beta=1$.\\
\begin{figure}
\centering
\includegraphics[scale=0.6]{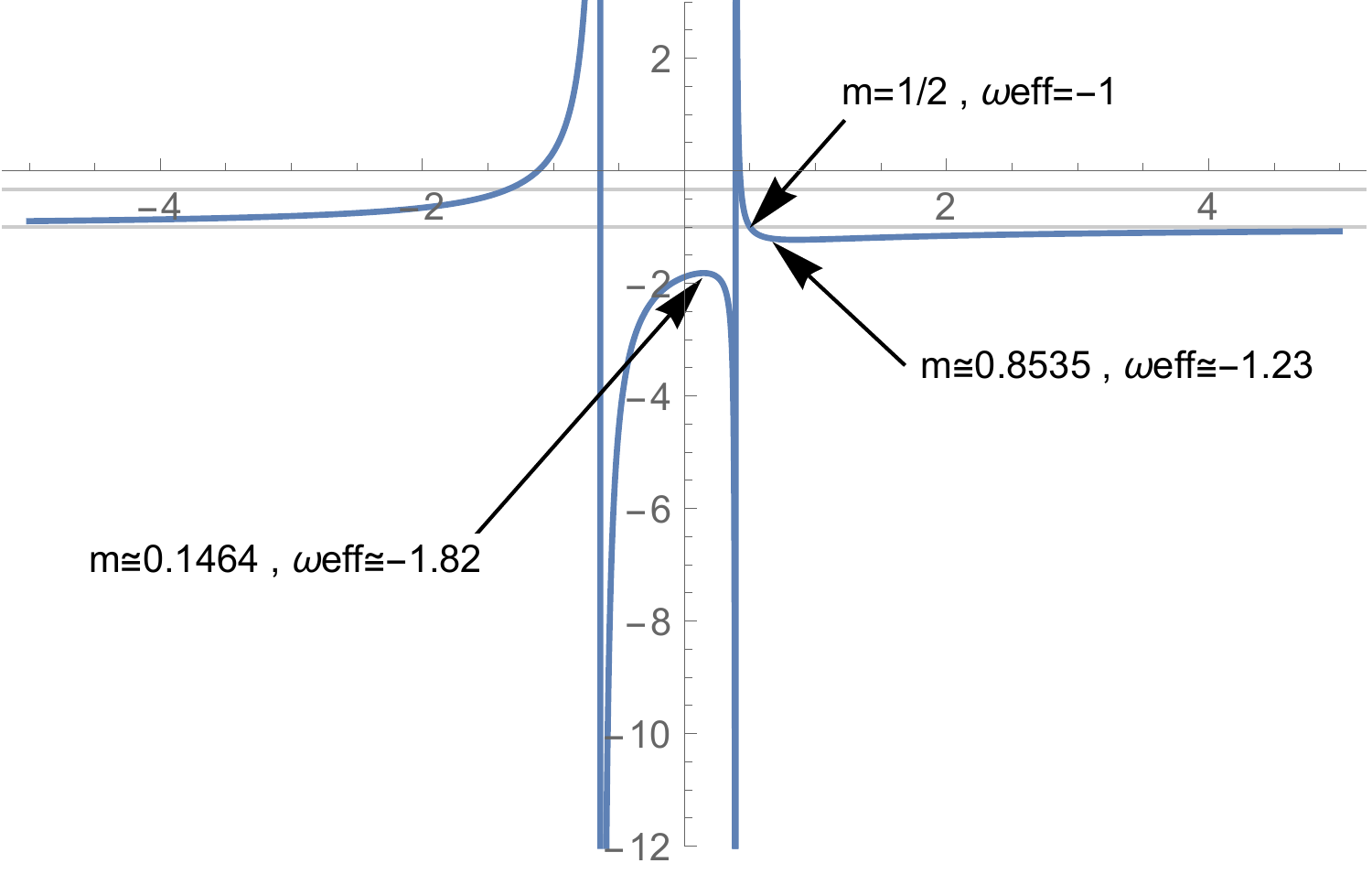}
\caption{The allowed regions for $w_{eff{P6}}$ in terms of $m$ including the holographic vacuum energy, with $\alpha=\frac{1}{3}$ and $\beta=\frac{1}{2}$.}
\end{figure}
\begin{figure}
\centering
\includegraphics[scale=0.6]{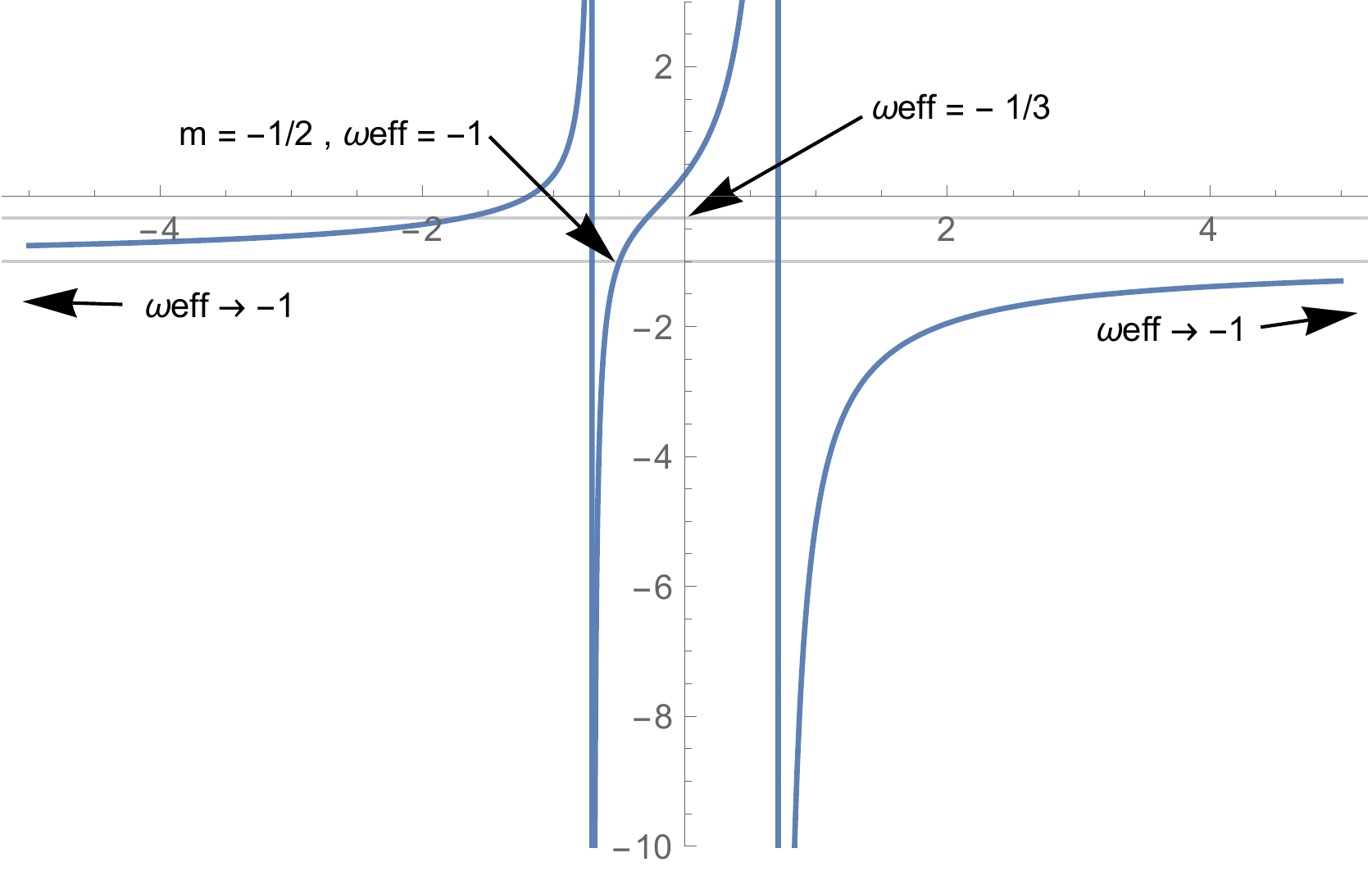}
\caption{The allowed regions for $w_{eff{P6}}$ in terms of $m$ including the holographic vacuum energy, with $\alpha=3$ and $\beta=1$.}
\end{figure}
\noindent As can be seen from Fig. 2 and 3, the $EoS$ in the framework of the holographic principle can take a wider range of values in the Phantom regime which did not happen with $f(R)$.\\
For the case of Fig. 1, if $m'<-1$ the point $\Tilde{P}_6$ is a saddle point in the Phantom regime for $-0.640388<m<0$ , and if $m'>-1$, $\Tilde{P}_6$ is an attractor for the same values of $m$. On the other hand, if $0<m<0.390388$, $\Tilde{P}_6$ is a saddle point for any value of $m'$.\\
For the case of Fig. 2, if $m'<-1$ all the Phantom region $\big(-\frac{1}{\sqrt{2}}<m<-\frac{1}{2}$ or $m>\frac{1}{\sqrt{2}}\big)$ of $P_6$ is transitory (saddle point) and if $m'>-1$ it is an attractor.\\
Let us go back to the de Sitter point $P_1$ and let us analyze the behaviour of it for  $\alpha=3$ and $\beta=1$. The second and third eigenvalues of $P_1$ (eqn. (\ref{EVP1})) are negative if  $-\frac{1}{2}<m<0$. It shows that in the frame of the holographic principle an $f(R)$ model can evolve from the transitory Phantom region of $P_6$ $\big(-\frac{1}{\sqrt{2}}<m<-\frac{1}{2}$ , $m>\frac{1}{\sqrt{2}}\big)$ with $m'<-1$ and $-\infty<w_{eff_{P_6}}<-1$ to the de Sitter attractor $P_1$ with $r=-\frac{1}{2}$ and $-\frac{1}{2}<m<0$ which is not possible with dark energy coming just from $f(R)$.\\
On the other hand, for the matter point $P_5$ it is possible to see that if $m\rightarrow 0$, $w_{eff_{P_5}}\rightarrow 0$, but $\Omega_{m_{P_5}}\rightarrow \frac{2-2\alpha+3\beta}{2}$, so in the holographic frame the matter content of $P_5$ depends on the holographic parameters $\alpha$ and $\beta$.\\
In the limit $m\rightarrow 0$, the eigenvalues of $\tilde{P}_5$ (without $\rho_\Lambda$) are: 
\begin{equation}
EV \tilde{P}_5 (\mid m \mid\ll1):\hspace{0.2cm}\Big\{3(1+m')\hspace{0.2cm},\hspace{0.2cm}-\frac{3}{4}\pm\sqrt{-\frac{1}{m}}\Big\}
\end{equation}

\noindent In this case, the second and third eigenvalues diverge as $m\rightarrow0^-$, one of them is negative and the other one positive. The positive divergence avoids the system to remain enough time in the matter dominated epoch so structure formation in the universe is difficult to take place except for a very narrow range of initial conditions. An approximated matter era can be achieved if $m<0$ ($m\nrightarrow0$, but close to zero), but the eigenvalues are large and it is difficult to find initial conditions to remain close to $\tilde{P}_5$ for a long time. Because this, in general, the $f(R)$ models (out of the holographic frame) with $m<0$ are not acceptable.\\
Otherwise, if $m\rightarrow0^-$ and $\beta>0$, the second eigenvalue of $P_5\rightarrow-\infty$ but the third eigenvalue can be modulated with the values of the holographic parameters $\alpha$ and $\beta$ (see eqn. (\ref{EVP5m0})). If $m'<-1$ and $\frac{2\alpha+\beta}{4}>\frac{3}{4}$, $P_5$ is a saddle point with one non-divergent positive eigenvalue giving place to a transitory point that can lead to a long enough matter era. The same situation occurs if $m'>-1$ no matter the values of $\alpha$ and $\beta$ as long as they do not diverge. If $m\rightarrow0^+$ and $\beta<0$, the second eigenvalue of $P_5$ is negative divergent and the achievement of a suitable (long enough) saddle matter point is driven by $m'$ ($m'>-1$ or $m'<-1$) in the first eigenvalue, and $\alpha$ and $\beta$ ($2\alpha-3>\mid\beta\mid$\hspace{0.2cm}or\hspace{0.2cm}$2\alpha-3<\mid\beta\mid$  respectively) in the third eigenvalue.\\ 
For these reasons, the inclusion of the holographic vacuum energy $\rho_\Lambda$ allows to cross $P_5$ with $m\rightarrow0^\pm$ which does not happen when the dark energy sector comes only from $f(R)$ \cite{Amendola2007}. For $P_5$ with $m\rightarrow0$, different matter dominated eras can be achieved:
\begin{equation}
    \begin{split}
\hspace{1.5cm}&\hspace{0.5cm}\Omega_{m_{{P}_5}}=1\hspace{0.2cm}\text{for}\hspace{0.2cm} \alpha=\frac{3}{2}\beta\hspace{0.5cm},\\&
\Omega_{m_{{P}_5}}=0.9\hspace{0.2cm}\text{for}\hspace{0.2cm} \alpha=\frac{3}{2}\beta+\frac{1}{10}\hspace{0.5cm},\\&
\hspace{0.1cm}\Omega_{m_{{P}_5}}=0.8\hspace{0.2cm}\text{for}\hspace{0.2cm} \alpha=\frac{3}{2}\beta+\frac{1}{5}\hspace{0.5cm},\\&
\Omega_{m_{{P}_5}}=0.7\hspace{0.2cm}\text{for}\hspace{0.2cm}\alpha=\frac{3}{2}\beta+\frac{3}{10}\hspace{0.3cm}.
    \end{split}
\end{equation}

\section{Results applied to the Hu and Sawicki's model}

In this successful model the $f(R)$ takes the form:

\begin{equation}f_{HS}(R)=R-m_s^2\frac{c_1\big(\frac{R}{m_s^2}\big)^n}{c_2\big(\frac{R}{m_s^2}\big)^n+1}\end{equation}

where $n>0$, $c_1$ and $c_2$ are dimensionless parameters, and $m_s$ is a mass scale given by $m_s\equiv\frac{\kappa^2\Bar{\rho_0}}{3}$ where $\Bar{\rho_0}$ is the average density today.\\
If $X\equiv\frac{R}{m_s^2}$, the $r$ and $m$ parameters are:
\begin{equation}r=\frac{c_1nX^n-X(1+c_2X^n)^2}{(1+c_2X^n)[X+(c_2X-c_1)X^n]}\end{equation}
\begin{equation}\text{and}\hspace{0.5cm}m=\frac{c_1nX^n(1-n+c_2(1+n)X^n)}{(1+c_2X^n)[X(1+c_2X^n)^2-c_1nX^n]}\hspace{0.5cm}.\end{equation}
Let us consider the following cases.\\

\noindent {\bf Example 1}: if $c_1=4$, $c_2=7$ and $n=5$, the trajectory in the $m$ vs $r$ plane is depicted in Fig. 4.

\begin{figure}
\centering
\includegraphics[scale=0.6]{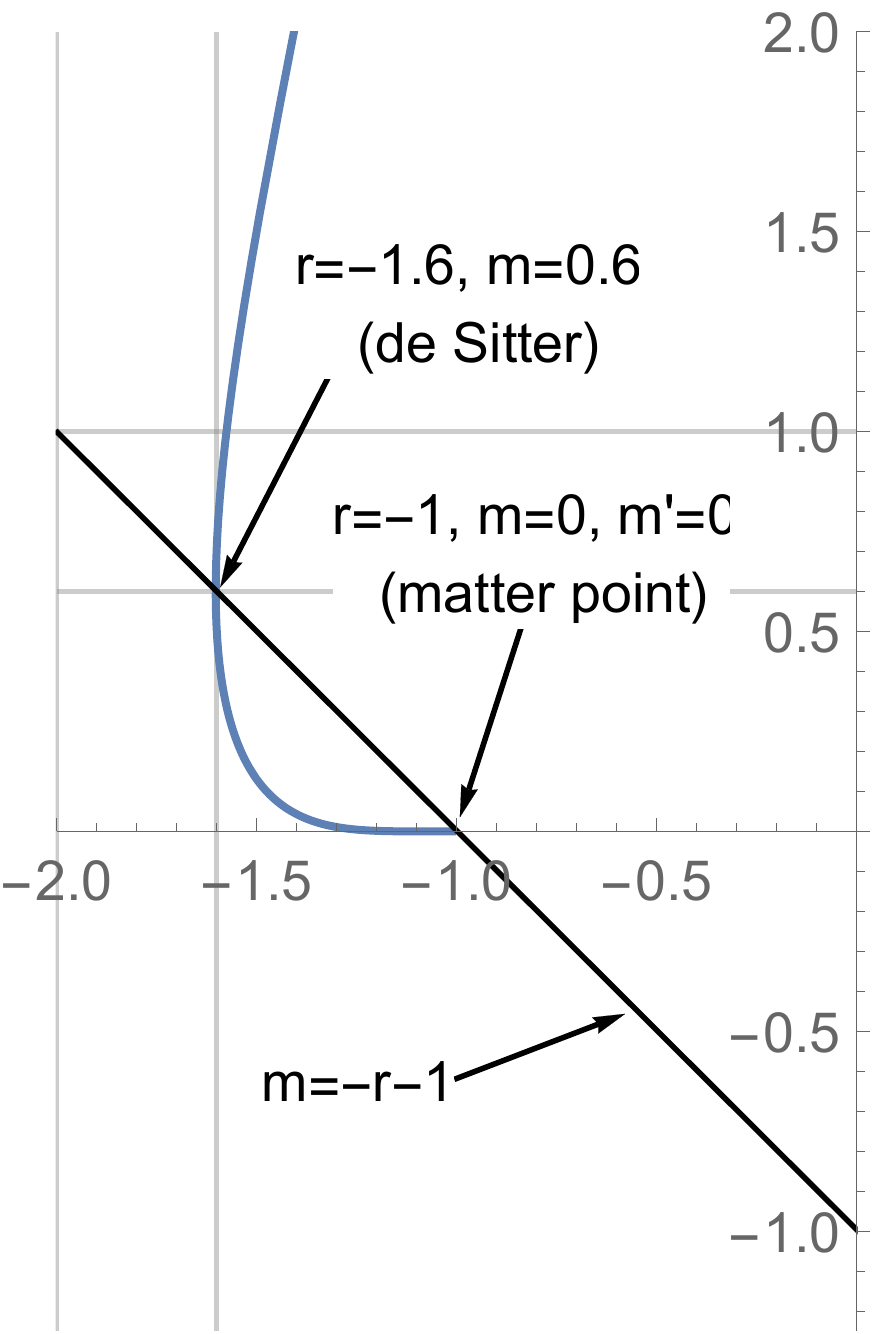}
\caption{$m$ vs $r$ for the Hu-Sawicki with $c_1=4$, $c_2=7$ and $n=5$ .}
\end{figure}

\noindent If $r\rightarrow-1^-$ then $m\rightarrow0^+$ and $m'\rightarrow0$, so if $\alpha=\frac{1}{4}$ and $\beta=-10^{-2}$, the eigenvalues in $P_5$ are $3$, $-\infty$ and $-0.627$, with $\Omega_{mP5}\approx 0.74$ and $w_{eff{P5}}\rightarrow0$, that corresponds to a saddle point associated to a matter dominated epoch that could last enough in order to let structure formation. With the same values of $\alpha$ and $\beta$, if $r\rightarrow-1.6^+$then $m\rightarrow0.6^-$, $\Omega_{mP6}=0$ and $w_{eff{P6}}=-1$. The associated eigenvalues are $0$, $-2.73$ and $-2.78$ which corresponds to a marginally  stable de Sitter point. This behaviour is not possible just with $f(R)$ where a de Sitter behavior can be found only when $r=-2$ for $\tilde{P}_1$, and $m=1$ or $m\rightarrow\pm\infty$ for $\tilde{P}_6$ .\\

\noindent {\bf Example 2}: If $c_1=3$, $c_2=15$ and $n=2$, gives the trajectory in the $m$ vs $r$ plane as shown in Fig.5.

\begin{figure}
\centering
\includegraphics[scale=0.4]{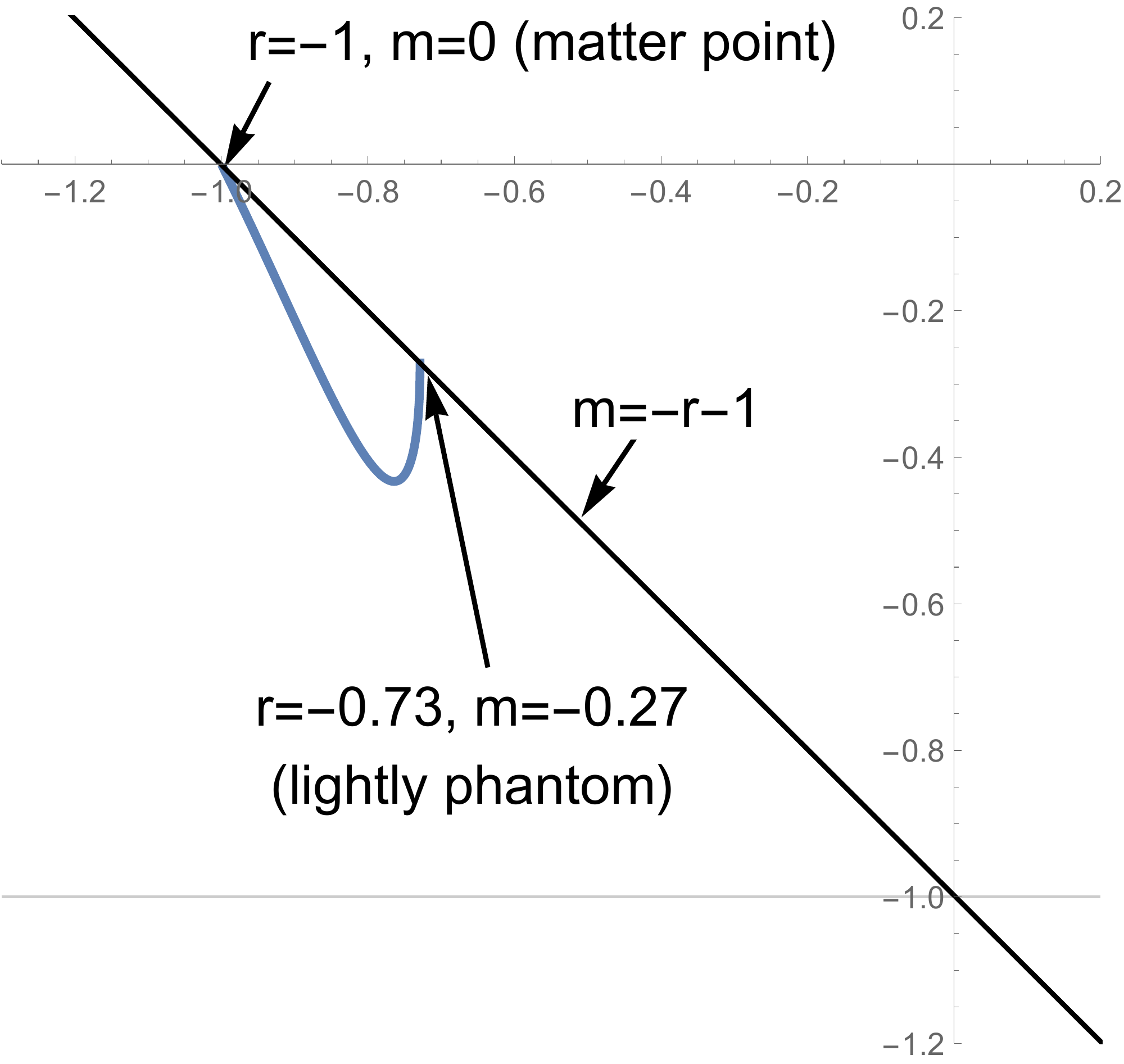}
\caption{$m$ vs $r$ for the Hu-Sawicki with $c_1=3$, $c_2=15$ and $n=2$.}
\end{figure}
\noindent If $r\rightarrow-1^+$, $m\rightarrow0^-$ and $m'<-1$, so if $\alpha=1.57$ and $\beta=1$, the first eigenvalue in $P_5$ is negative, and the other two are $-\infty$ and $0.285$ respectively, with  $\Omega_{m_{P_5}}=0.93$ and $w_{eff_{P_5}}=0$, corresponding to a saddle point of a matter dominated epoch that can be long enough for structure formation. With the same values of the holographic parameters, when $r\rightarrow-0.73^-$, $m\rightarrow-0.27^-$, and $\Omega_{m_{P_6}}=0$ and $w_{eff_{P_6}}=-1.1$. The eigenvalues in this point are $-\infty$, $-3.76$ and $-9.27$ so it is a lightly phantom attractor point. This behaviour is not possible with dark energy coming just from $f(R)$ because in that case when $m\rightarrow0^-$ the trajectories to $\tilde{P}_5$ are not allowed, and $w_{eff\Tilde{P}_6}=-1.1$ (as a limit of the current accelerated universe) is not reached.\\

\section{Discussion}

Under the assumption that modified gravity is not incompatible with the holographic principle, but rather may complement each other in the seek for the underlying theory of quantum gravity, we propose the study of modified gravity in the framework of the holographic principle.
As can be seen, the introduction of $\rho_\Lambda$ as source of vacuum energy, gives viability to different cosmological trajectories of $f(R)$ models which were not allowed out of the holographic frame. The critical points associated to matter domination and accelerated scenarios acquire new characteristics defined by the holographic parameters $\alpha$ and $\beta$ and their corresponding eigenvalues become more interesting since they give place to different stable or transitory cosmological solutions for a wider range of values of $r$, $m$ and $m'$, as shown in Figs. 2 and 3 for the general case and in Figs. 4, 5 for the particular case of the Hu-Sawicki model.\\
It would be interesting to perform further study of the perturbations of this model to analyze the structure formation in the frame of the holographic principle. 

\section*{Acknowledgments}
\noindent This work was supported by Universidad del Valle under project CI 71195.


\end{document}